# Towards a Secure Smart Grid Storage Communications Gateway

Stefan Marksteiner and Heribert Vallant

*Abstract*— This research in progress paper describes the role of cyber security measures undertaken in an ICT system for integrating electric storage technologies into the grid. To do so, it defines security requirements for a communications gateway and gives detailed information and hands-on configuration advice on node and communication line security, data storage, coping with backend M2M communications protocols and examines privacy issues. The presented research paves the road for developing secure smart energy communications devices that allow enhancing energy efficiency. The described measures are implemented in an actual gateway device within the HORIZON 2020 project STORY, which aims at developing new ways to use storage and demonstrating these on six different demonstration sites.

*Index Terms*—Cyber-physical systems, Energy storage, Information security, Power system management, Privacy, Smart grids

## I. INTRODUCTION AND MOTIVATION

IN this paper we propose security requirements and privacy recommendations for a smart grid storage communications gateway (SCG). The intended purpose of this gateway is to provide a means of communication to distribute information about flows between energy production and storage devices of various sizes. The gateway is developed within the HORIZON 2020 project *added value of STORage in distribution sYstems (STORY)*, in the context of ICT supporting services to enhance the efficiency of storage systems. Going to be dedicated to transmitting data and control signals within demonstration sites, this gateway is regarded critical infrastructure, which obviously has severe security implications. The objective of this paper is to clarify these implications for network nodes, communication lines, machine-to-machine (M2M) protocols, data storage and privacy aspects. The presented results are more developed, yet work in progress, research based on a first publication in 2016 [1].

Being designed for the communications gateway developed in the STORY project, they can, in principle, also be applied to any device with a similar profile (for instance other smart grid devices or smart appliances). For devices partially matching this profile (for instance ICT-enabled devices with well-known peers, see Section IV), the requirements concerning the matching parts may be applied.

## II. RELATED WORK

While there is some theoretical work [2,3] on security aspects in smart energy systems and solutions for distinct problems in this area available [4,5], there is no comprehensive hands-on configuration guidance for developers of such energy devices. Therefore, the recommendations and requirements in this paper will serve also as a practical reference for connecting and interoperating M2M networks to ICT networks in the context of energy production, distribution, consumption and storage devices. This applicability set this paper in contrast to known security standards in the energy context operating at this level of detail, which are mostly limited to smart metering, like the protection profile of the German Office for Information Security [6]. Standards that follow a more general approach, on the other hand, are too unspecific to be used for practical implementation and are mostly designed for overall systems architectures rather than providing device-level guidance [7].

Other known approaches have a different focus regarding device purpose and security focus. The OGEMA project, for instance, aims on an architecture more resembling the emerging STORY Smart Energy Platform (see Section VII) and focuses its security effort to application security and user rights management [8].

## III. STORY PROJECT OVERVIEW

STORY presents six different demonstration cases, each with different local/small-scale storage concepts and technologies, covering industrial and residential environments.

These cases are situated in the mid to western parts of Europe:
  1) Demonstration at residential building scale, Oud-Heverlee, Belgium
  2) Demonstration at residential neighborhood scale, Oud-Heverlee, Belgium
  3) Demonstration of storage in a factory, Navarra, Spain
  4) Demonstration of storage in residential district, Lecale, Northern Ireland

Manuscript received March 20, 2017; revised April 27, 2017
This work has received funding from the European Union's Horizon 2020 research and innovation programme under grant agreement No. 646426 Project STORY–H2020–LCE-2014-3.
The authors are with the Connected Computing Research Group, DIGITAL – Institute for Information and Communication Technologies, JOANNEUM RESEARCH, Graz, Austria. (e-mail: stefan.marksteiner@joanneum.at, heribert.vallant@joanneum.at).

5) Demonstration of large scale storage unit in industrial and residential area, Hagen, Germany and Suha, Slovenia
6) Demonstration of private multi energy grid in industrial area, Olen, Belgium

The management of these different STORY systems in a smart multi-energy grid is a crucial element to facilitate communication, integrate control algorithms, increase interoperability and provide measurement services.

### A. STORY ICT Architecture

Figure 1 displays the communication networks of STORY demonstration sites on a general level. The following three network types, based on their application location, are identified and for each type fixed, wireless, and mobile technologies can be applied:

- Home Area Network (HAN), also called a Premises Area Network or a Subscriber Access Network;

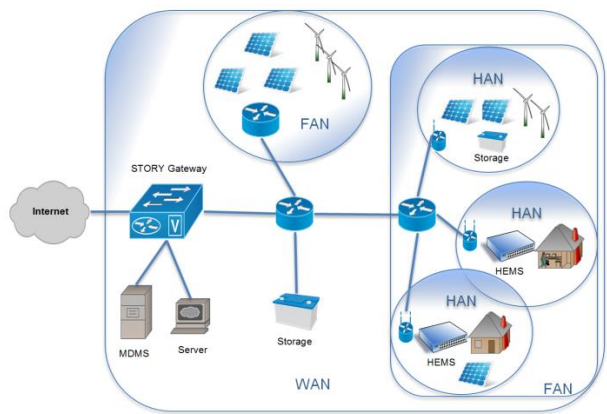

Fig. 1 Overview of communication networks in demonstration sites. [9]

- Field Area Network (FAN), also called a neighborhood Area Network (NAN);
- Wide Area Network (WAN).

Depending on the size of the demonstration site different combinations of these network layers have been deployed. The common entities for all demonstrations are the gateway device, energy storage device and a WAN network. This view of the STORY architecture is idealized and generic. During the course of the project, it became evident that the communication levels are different at each demo site. If and how this could be harmonized is subject to further investigations (see Section VII).

STORY demonstration sites will be connected to the Internet via the gateway device which is being developed within the project. All data (measurements, faults, warnings, acknowledgements and control signals) transferred within the system have to pass this SCG. The gateway therefore resembles a smart energy gateway, connecting prosumer systems to a power grid.

Also, to broker applications between application providers and end users (prosumers), a separate Smart Energy Platform (SEP) will be developed (see Section VII).

### IV. SECURITY CONSIDERATIONS FOR THE STORAGE COMMUNICATIONS GATEWAY

One major challenge for the management of the proposed architecture is that energy requests and sensed measurement data have to be reliably transferred both internally within the demo sites and externally via a network outside of the demo site operators' sphere of influence. This reliable transfer includes, besides availability and robustness, also additional security and privacy issues that arise when data is transferred through external networks. This section therefore defines the security requirements and covers the privacy aspects for the different facets of the gateway.

### A. Node Security Requirements

For monitoring and configuration purposes, user interfaces will be required. This can be ranging from simple status views (for instance an energy level overview for prosumers) to interfaces for sophisticated data analytics for energy distribution prediction. In order to secure these interfaces, authentication and access control is one of the most crucial elements to secure the proposed system infrastructure. Therefore, depending on the size of the storage production capacity and their integration into the grid, two categories are introduced:

- For demo sites with storage capacity dimensioned mainly for their own consumption, individual user login/ password combinations or eIDs are proposed;
- For demo sites that are mainly under control by a DSO, a more sophisticated authentication process including a combination of a certificate (see [4] and role/ID-based authentication and access control process is planned.

The latter (two-factor) authentication method needs more administrative effort (especially establishing a public key infrastructure), but at DSOs, the presence of the needed expertise and resources is assumed. Therefore, the maintenance of certificates is assumed pre-existing.

The roles defined in the authentication (and authorization) may also be used for alerting; the system must be capable of reporting certain safety and security relevant events to distinctive people or groups thereof. Reporting methods may include email and SMS. Furthermore, the system may contain a security dashboard, displaying recent events on a management interface. We also propose up to eight and at least four levels of alert classes (each respectively corresponding to one or consolidating two Syslog logging levels [10]) to categorize these events and assign them to roles or groups to be alerted. Access control also includes measurements to prevent unauthorized access. To achieve this, the device must be hardened, which means that all unneeded services, users and file permissions (especially for executables with shell-like capabilities) must be deactivated. Further, a custom-compiled, reduced kernel (for instance not containing unneeded but reportedly error-prone network protocols or user interfaces) is also recommended to provide a smaller target to adversaries. Moreover, strict patch management and a host-based firewall must be in place.

*B. Communication Line Security Requirements*

Also, communication channels have to be secured. Due the limited resources normally available at embedded systems, a hybrid cryptosystem approach is used which combines the security benefit of a public-key cryptosystem with the efficiency of a symmetric-key cryptosystem. During the communication establishment, the asymmetric key is used to exchange the symmetric key, which used afterwards for an efficient data throughput. Such a hybrid approach is part of various cipher suites and should be enabled.

In order to provide authentication and encryption, Transport Layer Security (TLS) must be used to provide communication security while traversing over foreign networks. The reasons for this implementation to choose TLS over alternatives (e.g. IPsec) are the following [11]:
- TLS is easier to integrate between different vendors;
- TLS needs less overhead;
- TLS allows quicker handshakes;
- TLS is easier to configure.

These arguments apply particularly to the proposed gateway, as it will also be installed in prosumers' households where knowledge needed to configure more complex protocols will not be present and the gateway hardware will only have limited resources. In constrained environments such as STORY demo sites that largely consume their own production themselves, Datagram Transport Layer Security (DTLS) might be used as a lightweight alternative. For the remainder of this document, all requirements regarding TLS apply to DTLS in the same manner. TLS is sometimes confounded with the Secure Sockets Layer (SSL) protocol, but in contrast to the former it is not an Internet Engineering Task Force (IETF) standard, although there is a specification for historic purposes [12]. This confusion occurs because TLS is the standardized version of SSL [13, p. 60] and has been further enhanced since then [14,15]. The various versions of SSL, however, have proven to be insecure and must not be used, not even as fallback methods [16]. We further propose to use the most recent version of TLS.

To fully utilize the security features of TLS, this protocol stack has to be configured carefully. That means that encryption and authentication measures allowed in the standard, but regarded unsafe by now must not be used. The Internet Engineering Task Force (IETF) published a guideline for the secure use of TLS [17]. The gateway must support and implementers must comply to these recommendations with the following additional security-enhancing constrictions:
- Symmetric cipher with at least 128 bits (see below);
- Must not support static key assignments (RSA and PSK);
- Must negotiate the most recent version of TLS (currently 1.2; exclusion of TLS 1.0 and 1.1);
- Must implement strict TLS;
- Must disable TLS-level compression;
- DH keys of at least 2048 bits or ECDH keys of at least 192 bits must be used;
- No anonymous suite must be used [15];
- Authentication must be mutual (see below).

Unless stated otherwise, these constraints consist of setting recommendations from the IETF guideline cited above to requirements. For simplicity, the allowed TLS cipher suites are restricted to the ones recommended in the document [17]:
- TLS_DHE_RSA_WITH_AES_128_GCM_SHA256
- TLS_ECDHE_RSA_WITH_AES_128_GCM_SHA256
- TLS_DHE_RSA_WITH_AES_256_GCM_SHA384
- TLS_ECDHE_RSA_WITH_AES_256_GCM_SHA384

One reason for changing the IETF recommendations to mandatory is that the gateway is regarded critical infrastructure and has therefore a higher demand of security than standard desktop machines or web-enabled devices. Another reason is that generally industrial and energy systems have longer product life cycles and therefore the time period the security measures have to prevail are also longer. Because of this, the key size is restricted to 128 bits or greater, for they are recommended beyond 2031 by the National Institute of Standards and Technology (NIST) [18]. Another related issue is the block size of symmetric encryption algorithms. Block sizes of 64 bits are generally not recommended [19] and therefore prohibited for the gateway. Of the encryption algorithms currently standardized in TLS [20], only AES, SEED, CAMELLIA and ARIA fulfill this requirement. Of these four, SEED does not occur in a cipher suite in CCM/GCM mode and does therefore practically not fulfill these requirements. AES is strongly recommended, as it is the most proliferated of these algorithms. CCM and GCM are generally recommended, because both of them offer a combined authentication and encryption algorithm. Also, the gateway specification mentioned above is defined within the project, allowing the requirements to be more strict compared to the general TLS specifications, which is designed for systems that have to cope with a broader range of peers that have to interoperate. It is further assumed that peering gateway devices are known to each other a priori and therefore are able to authenticate mutually, which might not be possible for webservers, which are usually interacting with thousands of clients. All of these reasons above both allow and require restricting the cryptographic configurations to a narrower set than in an ordinary or more generic ICT infrastructure environment.

*C. Data Storage Requirements*

In the same manner as in communication channels, data stored locally on the device has to be secured from unauthorized access. Apart from system access controls, this data has to be encrypted and integrity checked by the same algorithmic methods as communication lines. This is distinctively a requirement for security relevant data (explicitly logs that contain security events), which must be encrypted and integrity checked. Additionally, for privacy reasons, some sort of anonymization method (see Section V) has to be implemented, if personal data is to be processed. As it is a sensitive part, special focus on the key management is needed. A key derivation function that is deemed state of the art by current research must be used. A smart card-based key derivation function is recommended. To protect systems (i.e.

ICS) in contact with the gateway, some sort of filtering (ICS intrusion protection or anomaly detection system) is also recommended. The model for this secure data storage is yet to be elaborated. Apart from that, a model that stores only minimum data directly on the device could be used. Instead, all data could be sent to a secure, trusted server within the service operator's trusted zone could be used. This however, does not nullify the need for storage security, for there will always be stored data for authentication (credentials or certificates for communications authentication) and device caching purposes. The latter is strongly advised to prevent data loss in case of unforeseen device resets or outage, in which case data might not be transferred correctly and timely. To achieve appropriate storage security, we recommend using disk encryption (for instance, using *dm-crypt* on Linux-based systems).

*D. Intra-site (M2M) Communication Requirements*

Although, due to the heterogeneous nature of the connected energy production, energy consumption and storage devices, it is currently difficult to give advice on secure M2M communications. The minimum requirement should be a protocol that provides at least a similar security level as the IEC 62351 standard [21], although this standard does not necessarily assure end-to-end security [22]. Due to the latter, still cryptographic end-to-end node protection, as described above should be in place. As the devices, ordinarily only operate with limited system resources, using only the shorter of the recommended key lengths and relying on elliptic curve cryptography is advisable [23].

As other M2M protocols (like MTConnect) lay even less (if any at all) focus on security, we strongly recommend using OPC UA (standardized as IEC/TR 62541 [24]) with WS Secure Conversation as means of intra-site M2M communications. Further research enhancing security in this area is expected as the importance of M2M communications is on the rise. Generally, we recommend, as it is common practice, to segregate Industrial Control System (ICS) and Supervisory Control and Data Acquisition (SCADA) networks as much as possible, both among each other (if there is more than one) and from ICT networks, especially the Internet. Means of segregation include (but are not limited to), depending on possibility and needed security level, physical isolation (through lack of connection or data diodes), VLANs, network layer segregation, whitelisting and network firewalls. Following the defense in depth principle (also called layered security) [25], we suggest not relying on a single line of protection, but securing every part of the network graph (Figure 1) separately, for experience has shown that there is always a certain probability that network segregation is not a hundred percent effective (for instance because of a posteriori design changes). Apart from that, established techniques from traditional network security, like rate limiting and filtering should be used [5]. Further considerations are out of scope, as this paper defines only the gateway's security requirements.

## V. PRIVACY CONSIDERATIONS FOR THE STORAGE COMMUNICATIONS GATEWAY

Privacy aspects in STORY are not limited to confidentiality and access control. The sensors in use will generate a large amount of data and partly highly sensitive personal data about activities within the demonstration site. At residential building demonstrations, the connection to smart household appliances or smart home functionality has to be considered, because it might have a huge impact on the privacy of a person. The amount of personal data that might be collected this way can potentially deliver a lot of information about a person's behavior, location and actions, as well as health and finance status. In the area of industrial demo sites, the interconnection with other deployed systems may have serious impact regarding accessibility of confidential internal information (data protection) and processes. Therefore, measurements have to be undertaken to protect this information from unauthorized access.

*A. Standards and Initiatives with Privacy context*

Our privacy recommendations draw from a series of European and international efforts to safeguard personal (end consumers') sensitive data. Primarily, the European General Data Protection Regulation, being *ius congens* in the EU's legal sphere, serves as pivotal point for our privacy considerations.

*1) European General Data Protection Regulation*

The European Union unified data protection within its jurisdiction with a single law, the General Data Protection Regulation (GDPR) [26]. This new legislation covers the so-called *right to be forgotten* and related rights for customers (restriction, rectification and erasure of personal data), as well as the data processing companies' obligation to classify, control and monitor data, including means to protect them from unauthorized access, further increasing the need for the measures proposed in Section IV. Non-compliance to this regulation is punishable with penalties of up to four percent of the corporation's global profit or 20 million Euros. This outspokenly customer-friendly legislation, in conjunction with the complexity of the potentially harvested and processed data creates the necessity for a data privacy model (see below).

*2) Other Initiatives*

In order to achieve a comprehensive view on privacy and give appropriate recommendation for data collection and processing in the storage communications gateway, a collection of initiatives related to data privacy was sighted and aggregated, yielding the recommendations in the next section. In particular, the considered initiatives and documents include the following:

- The OECD *Privacy Framework* [27];
- The *Safe Harbor Privacy* Principles [28] and them being rendered to supply only insufficient privacy guarantees [29];
- The EU-U.S. *Privacy Shield* [30], including an opposing view by European Digital Rights (EDRi) for

- not going far enough to protect the EU citizens' rights [31];
- The *IPEN* initiative [32];
- The Online Trust Alliance' *IoT Trust Framework* [33];
- The *OWASP Top 10 Privacy Risk* Project [34].

*B. Privacy Recommendations for the Storage Communication Gateway*

After sighting the sources mentioned above, the following privacy requirements were derived for the Storage Communication Gateway:

- All personally identifiable and sensible data must be encrypted using state of the art encryption standards;
- Establish state of the art access control mechanism to all data;
- Specify the nature and purpose of processed personally identifiable and sensitive data types and attributes;
- If sensitive data is transferred outside the premises, only the part of the data that is reasonably useful for the functionality must be allowed to be transferred;
- If data is transferred outside the premise, personalized data has to be pseudonymized;
- If data is transferred outside operations, personalized data has to be anonymized;
- During the transfer, all data has to be encrypted by using current generally accepted state of the art security standards;
- In general, data must only be stored within storage devices located inside the EU;
- Collected data should not be shared with third party organizations;
- Specify the data storage duration;
- Provide information about policies, terms and conditions to the user;
- Provide information and control mechanisms for users to decline collection and initiate removal of personalized data.

These requirements must be underpinned by an anonymization model which sets on *t-closeness* [35]. The model description is too extensive to be covered in this requirements specification. If personal data is to be transferred to third parties for processing and providing analytical (*big data*) services, it has occur via a trusted secure broker, allowing no linkage between processed data and personalities of the originator.

*C. Availability*

Availability requirements to the gateway were elaborated through a survey, consisting of questionnaires, sent out to each of demo sites. This survey, constituting an aggregated operator-side assessment, yielded that the gateway has to be available 24 hours a day (24/7), more precisely it has to exhibit an uptime of 99.9% (*three nines*) for STORY operations. To achieve this objective, a backup solution for data and configuration information is required. In general, the gateway itself has to be linked to the power supply directly. Depending on the local operation mode at the STORY demo site, an islanded mode without power from the grid might be required. Considering this, a battery or uninterruptible power source (UPS) to enable such a mode is required to ensure both operational availability at such high level and data loss prevention.

## VI. CONCLUSION

This research in progress paper outlined the requirements for a Secure Smart Grid Storage Communications Gateway as defined by the needs of the demonstration cases of the STORY project. The key findings are as follows:

- Node security requires an access control (ideally cryptographically supported) and monitoring concept, as well as device hardening;
- Communication channels should be secured with a recent TLS implementation using an AES cipher with 128 bit keys or more, an ephemeral key exchange algorithm, SHA2 hashing with 256 bits or more and the use of the GCM cipher mode;
- Data storage needs a secure access and anonymization concept, combined with storage-level encryption;
- Intra-site communications should use the OPC UA protocol (and make use of its security features) and layered defenses;
- The availability of the gateway should at least reach 99.9% (*three-nines*) using appropriate power redundancy and UPS concepts;
- A privacy concept based on *t-closeness* has, in conjunction with data storage, to be developed, coping with access protection and anonymization including their legal issues;

These findings allow building a secure smart grid storage communications gateway, as well as other building blocks to enhance energy efficiency in a way that is secure from an ICT perspective. As this is research in progress, the next section outlines identified potential for further research.

## VII. FURTHER RESEARCH

As this paper outlines the requirements for a secure smart grid storage communications gateway, the next logical step is its implementation. Therefore, further research is needed in technology and procedures (some of which have yet to be developed) that meet those requirements. In order to meet the privacy recommendations outlined in this document, also a statutorily regulation and technological model (including the secure storage of data both on the device and at different sites) has to be developed. Further, as the communication levels have proven to be different at each demo site, the implications of this circumstance on the security requirements have to be investigated.

To enhance the overall value of the project, the project team is also dedicated to develop a standalone Smart Energy Platform (SEP). This platform basically allows brokering the services of application providers to end users, as well as the use of data to generate billing and energy data statistics. Figure 2 shows the Smart Energy Platform within the STORY context. The security requirements are yet to be elaborated, although it is by now advised to lead all communication lines

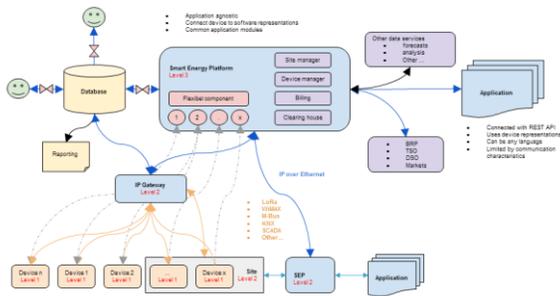

Fig. 2 Overview of the Smart Energy Platform architecture and context

through the SEP and the gateway to both minimize attack vectors and reduce interoperability issues.

ACKNOWLEDGMENT

The authors wish to express their gratitude to the STORY demonstration sites and the WP4 team, especially to Arnout Aertgeerts, Timo Kyntäjä, Paul Valckenaers, Topi Mikkola and Franci Katrašnik for their valuable inputs on the security considerations outlined in this paper.